\documentclass[
  aps,
  prl,
  reprint,
  10pt,
  superscriptaddress,
  longbibliography
]{revtex4-2}

\usepackage{amsmath,amssymb}
\usepackage{booktabs}
\usepackage{dsfont}
\usepackage{braket}
\usepackage{tikz}
\usetikzlibrary{calc,decorations.markings}
\usepackage[colorlinks=true,allcolors=blue]{hyperref}

\newcommand{\ee}{\mathrm{e}}
\newcommand{\ii}{\mathrm{i}}
\newcommand{\TT}{\mathcal{T}}
\newcommand{\GG}{\mathcal{G}}
\newcommand{\Tr}{\mathrm{Tr}}
\newcommand{\Id}{\mathds{1}}
\newcommand{\abs}[1]{\lvert#1\rvert}
\newcommand{\ccc}{\mathfrak{c}_{-}}

\newcommand{\calZ}{\mathcal{Z}}
\newcommand{\calV}{\mathcal{V}}

\makeatletter
\def\maketitle{
\@author@finish
\title@column\titleblock@produce
\suppressfloats[t]
}
\makeatother

\makeatletter

\newcommand*{\wideboxed}[1]{\setlength{\fboxsep}{1ex}%
  \fbox{\m@th$\displaystyle#1$}}
\makeatother
\begin{document}

\title{Chiral Central Charge from Real-Space Twist Operator Correlator}

\author{Yunlong Zang}
\email{zangyunlong22@mails.ucas.ac.cn}
\affiliation{Kavli Institute for Theoretical Sciences and School of Quantum, University of Chinese Academy of Sciences, Beijing 100190, China}

\date{\today}

\begin{abstract}
Recent years have witnessed substantial progress in the entanglement-based characterization of quantum phases of matter, with growing interest in real-space formulas due to their theoretical and experimental relevance. 
For two-dimensional gapped quantum systems, we propose a universal real-space formula for the chiral central charge based on the twist-operator correlator $Z_{g,h}=\langle\psi^{(N)}\rvert\mathcal{T}_A(g)\mathcal{T}_B(h)\mathcal{T}_C(g)\mathcal{T}_C(h)\lvert\psi^{(N)}\rangle$, where $\lvert\psi^{(N)}\rangle=\lvert\psi\rangle^{\otimes N}$ is the $N$-copy state and $g,h\in S_N$ are permutations. 
The twist operator $\mathcal{T}_R(g)$ permutes the $N$ replicas within region $R$ according to $g$. For pairs $(g,h)$ satisfying the spherical condition, we show that the phase of the correlator is determined by the chiral central charge $\mathfrak{c}_{-}$ through $Z_{g,h}/\left\lvert Z_{g,h}\right\rvert=(\theta_g\theta_h/\theta_{gh})^{\mathfrak{c}_{-}}$, where $\theta_g^{\mathfrak{c}_{-}}$ is the topological spin of the twist defect labeled by $g\in S_N$. The construction applies to both bosonic and fermionic systems. We derive the formula within a field-theoretic framework under suitable assumptions and further test it numerically in microscopic lattice models. Our results establish a universal connection between real-space entanglement structures and the chiral central charge, providing a systematic framework for characterizing chiral topological phases directly on the lattice.
\end{abstract}

\maketitle

\textbf{Introduction.---}
Over the past two decades, remarkable progress has been made in understanding quantum phases of matter~\cite{Wen2017Zoo,Zeng2019QIMatter}. In particular, two-dimensional gapped systems admit a long-wavelength description in terms of topological quantum field theory (TQFT), which organizes their classification and universal observables~\cite{Witten1989Jones,Kapustin2014SPT,Kapustin2015Fermionic}.

Such a field theory describes a renormalization-group fixed point, where nonuniversal ultraviolet details are absent. By contrast, a generic wave function away from a fixed point~\cite{chenguwen2010} has a finite correlation length and retains microscopic information. This raises a natural question: how can one extract universal topological invariants from a generic gapped many-body wave function in two dimensions? Real-space formulas~\cite{Shiozaki2017,Shiozaki2018,Cian2021,Turzillo2025partial,sohal2026symmetry} address this question by showing that topological properties are sufficiently robust to be detected directly from lattice degrees of freedom and remain well defined throughout a phase.

Entanglement~\cite{SHI2020,KitaevLocal} has made this extraction possible for an increasing range of topological data. Topological entanglement entropy detects the total quantum dimension~\cite{Kitaev2006TEE,Levin2006detecting}, while other real-space constructions probe fusion rules and topological spins~\cite{SHI2020,KitaevLocal,liu2024anyon,h53w-64dy}. These formulas clarify how continuum topological data emerge from microscopic degrees of freedom and can also provide practical diagnostics for numerical many-body states.

Here we focus on the chiral central charge $\ccc$, which characterizes the bulk thermal response and edge gravitational anomaly~\cite{KaneFisher1997Thermal}. Existing wave-function diagnostics~\cite{Tu2013Momentum} include the modular commutator and its R\'{e}nyi version~\cite{Kim2022chiral,Kim2022modular,Fan2022from,Fan2023Extracting,gass2025renyilike,sheffer2025probing}, which are formulated in terms of density matrices or entanglement Hamiltonians. The R\'{e}nyi modular commutator also admits an equivalent replica-space representation.
We present a generalization of this real-space formula to a broader class of replica-space permutations. The resulting phase depends only on the topological spins of twist defects and thereby probes chiral topological states.

\begin{figure}[t]
  \centering
 \includegraphics{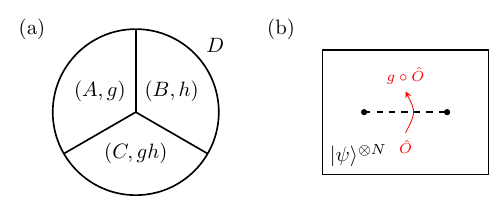}
  \caption{Twist-operator construction. (a) A large disk $A\cup B\cup C$ is divided into three regions, and $D$ denotes its complement. The labels specify the permutations acting in each region. (b) A twist operator supported on a closed region can be deformed to its boundary and then cut open into a twist-defect line. The endpoints (black dots) are twist defects, and the dashed line is the associated branch cut. Depending on the orientation convention, one endpoint is a $g$ twist defect and the other is a $g^{-1}$ twist defect. Crossing the cut permutes the replicas by $g$.}
  \label{fig:setup}
\end{figure}

\textbf{Main result.---}
Let $\lvert\psi\rangle$ be a gapped ground-state wave function in two spatial dimensions, and consider $N$ identical copies,
$\lvert\psi^{(N)}\rangle\equiv\lvert\psi\rangle^{\otimes N}$. As illustrated in Fig.~\ref{fig:setup}(a), divide a disk-shaped subregion into three adjacent parts $A$, $B$, and $C$, with all relevant length scales much larger than the correlation length. For $g,h\in S_N$, define
\begin{equation}
  Z_{g,h}
  =\bra{\psi^{(N)}}
  \TT_A(g)\TT_B(h)\TT_C(g)\TT_C(h)
  \ket{\psi^{(N)}} .
  \label{eq:def_Z}
\end{equation}
Here $\TT_R(g)$ is the \textit{twist operator}, which permutes the replicas only within region $R$, where $R\in\{A,B,C,D\}$. For a pair $(g,h)$ satisfying the \textit{spherical condition} defined below, we find that the phase of $Z_{g,h}$ detects the chiral central charge:
\begin{equation}
   \wideboxed{
  \frac{Z_{g,h}}{\abs{Z_{g,h}}}
  =\left(\frac{\theta_g\theta_h}{\theta_{gh}}\right)^{\ccc} .
   }
  \label{eq:phase_Z}
\end{equation}
The right-hand side depends only on $\ccc$ and the cycle structures of $g$, $h$, and $gh$; microscopic contributions affect the magnitude of $Z_{g,h}$ but not the phase in Eq.~\eqref{eq:phase_Z}.
Below, we explain Eqs.~\eqref{eq:def_Z} and \eqref{eq:phase_Z} in terms of three ingredients: the twist-defect spin, the spherical condition, and the definitions of bosonic and fermionic twist operators.

\emph{(i) Twist-defect spin.---}
Write the disjoint-cycle decomposition of $g$ as $g=\prod_{c\subset g}c$, including trivial one-cycles, and let $\abs{c}$ denote the length of $c$. The twist-defect spin at unit chiral central charge~\cite{Calabrese_2004} is
\begin{equation}
  \theta_g
  =\exp\!\left[-
    \frac{2\pi\ii}{24}
    \sum_{c\subset g}
    \left(\abs{c}-\frac{1}{\abs{c}}\right)
  \right].
  \label{eq:defect-spin}
\end{equation}
The corresponding twist defect of a gapped state with chiral central charge $\ccc$ therefore has topological spin $\theta_g^{\ccc}$~\footnote{Our convention may include an additional minus sign in the definition of $\theta_g$ compared to others.}. 
A twist defect is the analogue of an Aharonov--Bohm flux associated with the replica symmetry $S_N$~\cite{Barkeshli2013Genons,Barkeshli2019SFDG}. As shown in Fig.~\ref{fig:setup}(b), a $g$ branch cut connects two oppositely oriented pointlike twist defects, one labeled by $g$ and the other by $g^{-1}$. Crossing this branch cut implements the action of $g$. Equation~\eqref{eq:phase_Z} combines the contributions from the topological spins of the $g$, $h$, and $gh$ twist defects.
The importance of the twist-defect spin as a probe of the chiral central charge was emphasized by Sopenko~\cite{sopenko2024_1,sopenko2025_2}.

\emph{(ii) Spherical condition.---}
Define the transposition length $\ell(g)$ as
\begin{equation}
  \ell(g)\equiv\sum_{c\subset g}(\abs{c}-1)=N-n_{\mathrm{cyc}}(g),\quad g\in S_N.
  \label{eq:length}
\end{equation}
Here $n_{\mathrm{cyc}}(g)$ is the number of cycles of $g$, including one-cycles.
The spherical condition is
\begin{equation}
  \ell(g)+\ell(h)=\ell(gh).
  \label{eq:spherical}
\end{equation}
Geometrically, the spherical condition ensures that the path-integral representation of $Z_{g,h}$ corresponds to a three-sphere $S^3$ (or a disjoint union of several three-spheres), hence the name~\cite{SM}.

\emph{(iii) Bosonic and fermionic twists.---}
We now specify the explicit form of twist operators on the lattice.
For bosonic degrees of freedom, the on-site twist operators form an ordinary representation of $S_N$:
\begin{equation}
  \TT_{\mathrm b,\boldsymbol{x}}(g)
  \hat b_{\boldsymbol{x}}^{(r)}
  \TT_{\mathrm b,\boldsymbol{x}}^{-1}(g)
  =\hat b_{\boldsymbol{x}}^{(g[r])} .
  \label{eq:bosonic-twist}
\end{equation}
Here $r\in\{1,2,\dots,N\}$ is the replica index, and $g[r]$ denotes the action of $g$ on $r$.
Fermionic permutations require a sign convention~\cite{Casini_2005}. For a cycle $c=(r_1r_2\cdots r_M)$ of $g$,\footnote{To avoid ambiguity, we choose $r_1$ to be the smallest element of the cycle $c$.} we choose
\begin{equation}
  \TT_{\mathrm f,\boldsymbol{x}}(g)
  \hat f_{\boldsymbol{x}}^{(r_j)}
  \TT_{\mathrm f,\boldsymbol{x}}^{-1}(g)
  =
  \begin{cases}
    -\hat f_{\boldsymbol{x}}^{(g[r_j])}, & j\ne M,\\
    \phantom{-}\hat f_{\boldsymbol{x}}^{(g[r_j])}, & j=M.
  \end{cases}.
  \label{eq:fermionic-twist}
\end{equation}
Unlike their bosonic counterparts, the fermionic twists generally satisfy
$\TT_{\mathrm f}(g)\TT_{\mathrm f}(h)\neq\TT_{\mathrm f}(gh)$, so they are twist operators~\footnote{We use this terminology because such operators appear in calculations of fermionic entanglement entropy.} rather than permutation operators. This is also why Eq.~\eqref{eq:def_Z} retains the ordered product~\footnote{We choose this ordering because our formula must reduce to the R\'{e}nyi modular commutator in the simplest case.} $\TT_C(g)\TT_C(h)$ instead of replacing it by $\TT_C(gh)$.
In summary, the twist operator in region $R$ is $\TT_R(g)=\prod_{\boldsymbol{x}\in R}\TT_{\mathrm b/\mathrm f,\boldsymbol{x}}(g)$, and the same formula applies to both bosonic and fermionic systems.

Because no general proof exists for interacting systems, Eq.~\eqref{eq:phase_Z} remains conjectural, as does the modular commutator. Below, we give a field-theoretic derivation under suitable assumptions, followed by numerical lattice tests; further details are provided in the Supplemental Material~\cite{SM}.

\textbf{Field-theoretic derivation.---}
\begin{figure}[t]
\centering
  \includegraphics{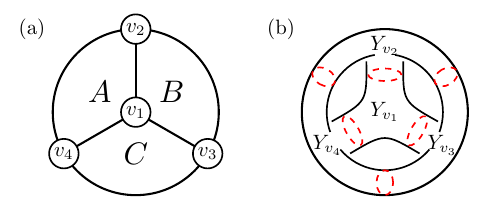}
\caption{(a) Tripartite region $ABC$ with the four triple points $v_1,v_2,v_3,v_4$. The black lines are the entanglement cuts of this tripartite geometry. (b) The genus-three entangling surface $Y$ is further cut into four three-punctured spheres $Y_v$ ($v=v_1,v_2,v_3,v_4$).}
\label{fig:def_Y}
\end{figure}
We interpret $Z_{g,h}$ as the partition function of a TQFT on a three-manifold obtained by wave-function surgery dictated by the permutations. As pointed out in Ref.~\cite{sheffer2025probing}, systems with a nonvanishing chiral central charge $\ccc$ require an additional regularization of the entanglement cuts---namely, the introduction of a chiral conformal field theory (CFT)---which is essential for obtaining the correct phase of the partition function. The entanglement cut is thickened into a tubular region whose boundary is the genus-three entangling surface ${Y}$ [Fig.~\ref{fig:def_Y}(b)]. In replica space, $N$ copies of $Y$ are cut and reglued according to the permutation rules to form a closed two-manifold $\Sigma$. When omitting framing anomaly~\cite{Witten1989Jones}, the phase of the partition function is then determined entirely by the physics on $\Sigma$:
\begin{equation}
    \arg Z_{g,h} = \arg \mathcal{Z}_{{\rm CFT}}(\Sigma),\quad 
\Sigma=\left(\bigsqcup_{r=1}^{N}Y^{(r)}\right)\Big/\sim_{g_A,g_B,g_C}.
\end{equation}
The evaluation of $Z_{g,h}$ thus reduces to computing the two-dimensional CFT partition function on $\Sigma$, whose geometry and connectivity are dictated by the permutations $g_A=g$, $g_B=h$, and $g_C=gh$ acting in regions $A$, $B$, and $C$, respectively.

\begin{figure}[t]
\includegraphics{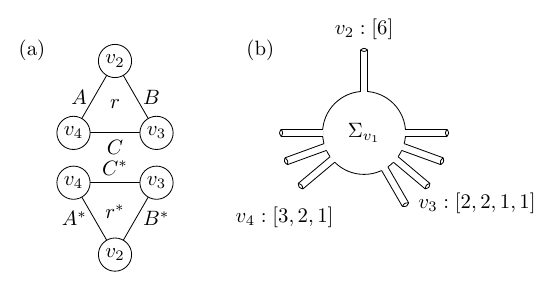}
\caption{(a) The replica $Y_{v_1}^{(r)}$ is cut into bra and ket sheets, where the replica index is $r=1,2,\dots,N$. (b) For the example permutations in Eq.~\eqref{eq:exampleS6}, $\Sigma_{v_1}$ is an eight-punctured sphere. The eight punctures are divided into three groups given by the preimages of $v_2$, $v_3$, and $v_4$, which are connected to $\Sigma_{v_2}$, $\Sigma_{v_3}$, and $\Sigma_{v_4}$ by long cylinders. The ramification indices are specified by the cycle types of the relative permutations $[g_{AB}]$, $[g_{BC}]$, and $[g_{AC}]$.}
\label{fig:exampleS6}
\end{figure}
\begin{figure}[t]
  \centering 
  \includegraphics{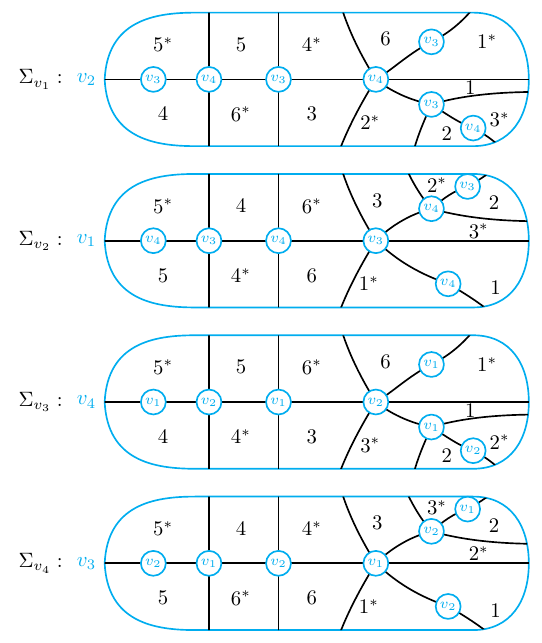}
  \caption{The four surfaces $\Sigma_v$ for $g_A=(23)(46)$, $g_B=(136)(45)$, and $g_C=(123456)$, with cycle counts $4$, $3$, and $1$, respectively. The relative permutations determine the ramification indices and have cycle types $[g_{AB}]=[g_{CD}]=[6]$, $[g_{BC}]=[g_{AD}]=[2,2,1,1]$, and $[g_{AC}]=[g_{BD}]=[3,2,1]$.}
  \label{fig:S6sigmav}
\end{figure}

Following Ref.~\cite{sheffer2025probing}, we cut the entangling surface ${Y}$ into four three-punctured spheres $Y_{v}$ ($v=v_1,v_2,v_3,v_4$), as shown in Fig.~\ref{fig:def_Y}(b). Gluing $N$ copies of each $Y_v$ produces four surfaces $\Sigma_v$. This first stage of the construction does not contribute to the complex phase. The phase arises only when the four surfaces $\Sigma_v$ are glued together by connecting their respective punctures to form $\Sigma=\left(\bigsqcup_v \Sigma_v \right)/\sim$.

We illustrate the above process with a concrete example in $S_6$:
\begin{equation}
    \begin{aligned}
        &g_A = (23)(46),\\
        &g_B = (136)(45),\\
        &g_C = (123456)=g_A g_B.
    \end{aligned}
    \label{eq:exampleS6}
\end{equation}
To construct $\Sigma_{v_1}$, we cut each replica $Y_{v_1}^{(r)}$ into a bra ($r^*$) sheet and a ket ($r$) sheet [Fig.~\ref{fig:exampleS6}(a)]. We then glue the six replicas according to the permutations to form $\Sigma_{v_1}$, which in this case is an eight-punctured sphere [Fig.~\ref{fig:exampleS6}(b)]. Both $\Sigma_{v_1}$ and $Y_{v_1}$ are Riemann surfaces, and the covering map $\Sigma_{v_1}\to Y_{v_1}$ is $N$-sheeted and locally one-to-one except at finitely many points.
Local injectivity fails only at the punctures, where the map becomes locally many-to-one. Such a puncture is a ramification point, at which the map locally behaves as $z\mapsto z^k$, with $k$ the \textit{ramification index}. In Fig.~\ref{fig:exampleS6}(b), the punctures are divided into three groups given by the preimages of $v_2$, $v_3$, and $v_4$. The number of punctures in each group equals the number of cycles of the corresponding relative permutation, and the ramification indices are the associated cycle lengths, namely, the entries in its cycle type. For example, $v_4$ has three preimages with ramification indices $[g_A g_C^{-1}]=[3,2,1]$.
The same procedure yields the remaining surfaces $\Sigma_v$ and the ramification indices of their punctures.

The next step is to pair and connect punctures on the four surfaces $\Sigma_v$. Because the punctures are small, connecting a pair of them produces a long, thin cylinder, and propagation along the cylinder projects the CFT onto its ground state. A ramification point with $k>1$ additionally induces a geometric twist angle $\tau_k$ along the cylinder. Each cylinder therefore contributes to the phase as
\begin{equation}
  \bra{0}\ee^{\ii\tau_{k} \hat{P}_{\textrm{CFT}}}\ket{0}=\ee^{-\frac{\ii\ccc}{24}\tau_{k}},
\end{equation}
where $\hat{P}_{\textrm{CFT}}$ is the CFT momentum operator. It remains to determine the twist angle of each cylinder.

For example, in Fig.~\ref{fig:S6sigmav}, the cylinder formed by gluing the $v_3$ puncture of $\Sigma_{v_1}$ to the $v_1$ puncture of $\Sigma_{v_3}$ is bordered by the four sectors $|5|4^*|3|6^*|$ and therefore has ramification index $2$. In general, a ramification index $k$ implies $2k$ sectors, as shown in Fig.~\ref{fig:S6sigmav}. In the Supplemental Material~\cite{SM}, we show that the corresponding geometric twist angle is
\nopagebreak[4]
\begin{equation}
  \tau_k =\left\{
    \begin{aligned}
      &\pi\frac{k-1}{k},\quad k<N, \\
      &-\pi\frac{N-1}{N},\quad k=N.
    \end{aligned}
  \right.
  \label{eq:tauk}
\end{equation}
The minus sign arises because each $\Sigma_v$ has a unique ramification-$N$ puncture at infinity, as shown in Fig.~\ref{fig:S6sigmav}. This puncture contributes with the opposite sign relative to all punctures of lower ramification index. In particular, the spherical condition ensures that
\begin{equation}
  [g_{AB}]=[g_{CD}]=[N],
\end{equation}
where $g_{IJ}=g_I g_J^{-1}$. Furthermore, the remaining ramification indices are determined by the cycle types of the relative permutations:
\begin{equation}
  \begin{aligned}
    &[g_{BC}]=[g_{AD}]=[g_A]\equiv [g],\\
    &[g_{AC}]=[g_{BD}]=[g_B]\equiv [h].
  \end{aligned}
\end{equation}
The cycle types of the relative permutations therefore suffice to determine all the twist angles. Summing the contributions from all cylinders gives
\begin{equation}
  \begin{aligned}
    \arg Z_{g,h}&= -\frac{\ccc}{24} \sum_{IJ} \tau_{IJ}\\
    &= -\frac{2\pi}{24}\ccc \left[\left(\sum_{c\subset g,h}\left(1-\frac{1}{\abs{c}}\right)\right)-\left(1-\frac{1}{N}\right) \right] \\
    &=-\frac{2\pi}{24}\ccc \left[\left(\sum_{c\subset g,h}\left(\abs{c}-\frac{1}{\abs{c}}\right)\right)-\left(N-\frac{1}{N}\right) \right],
  \end{aligned}
\end{equation}
where $IJ\in\{AB,AC,BC,AD,BD,CD\}$. The second equality uses Eq.~\eqref{eq:tauk} with $k$ set by the cycle lengths of the relative permutations $g_{IJ}$, while the last equality follows from the spherical condition, completing the field-theoretic derivation of our main formula.

\textbf{Numerics.---}
\begin{figure}
    \includegraphics[width=\columnwidth]{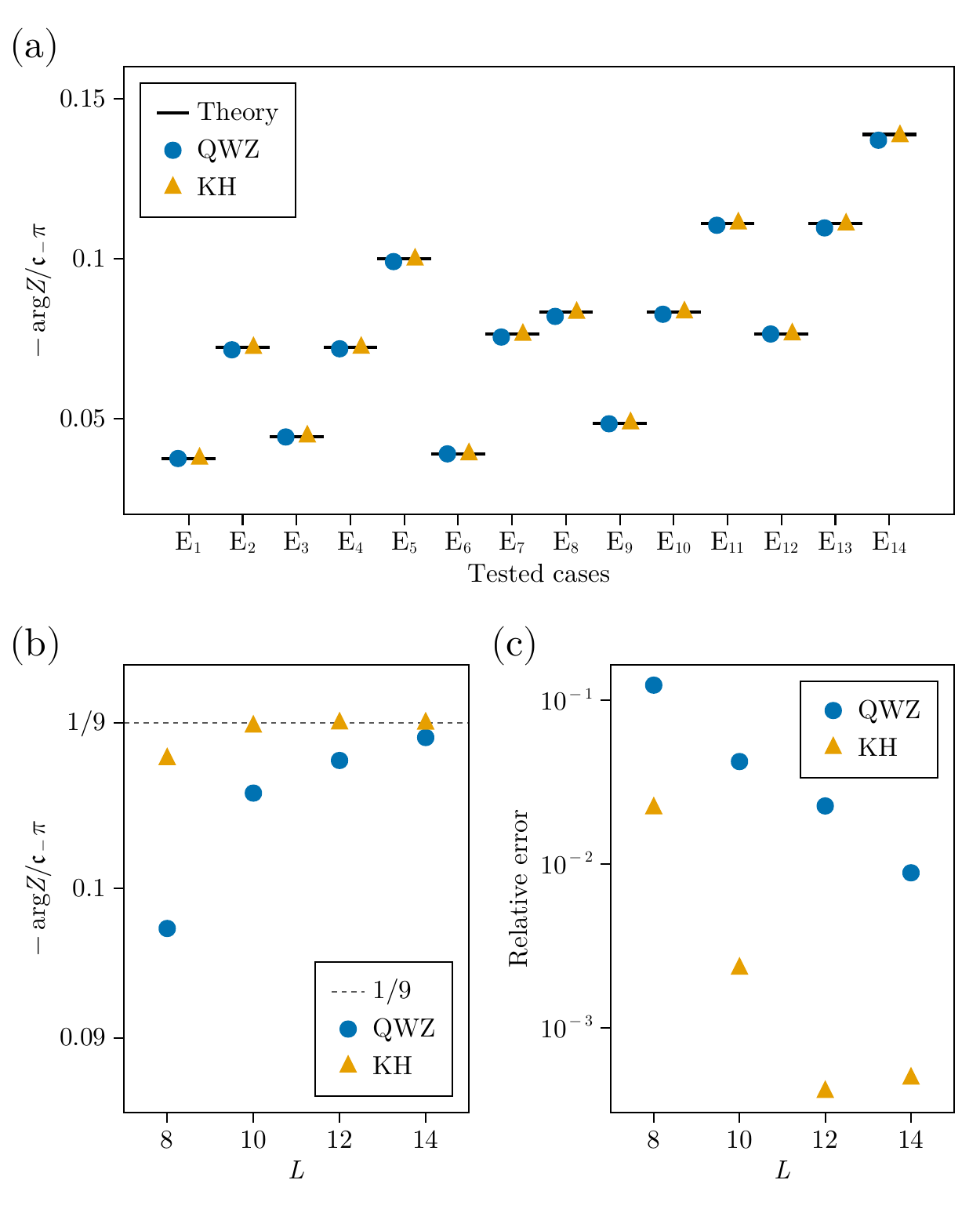}
    \caption{Numerical test of Eq.~\eqref{eq:phase_Z}.
    (a) $-\arg Z_{g,h}/(\ccc\pi)$ for 14 spherical pairs $(g,h)=(g_A,g_B)$, five in $S_5$ ($E_1$--$E_5$) and nine in $S_6$ ($E_6$--$E_{14}$), computed in the QWZ (circles) and KH (triangles) models; solid lines mark the theoretical prediction. The explicit permutations $(g_A,g_B,g_C)$ of each case are listed in the Supplemental Material~\cite{SM}.
    (b) Finite-size behavior of $-\arg Z_{g,h}/(\ccc\pi)$ for the $S_6$ example of Eq.~\eqref{eq:exampleS6}, obtained on $2L^2$ lattice sites with $L=8,10,12,14$; the dashed line marks the predicted value $1/9$.
    (c) Relative error of the data in (b) with respect to $1/9$.
    }
    \label{fig:num}
\end{figure}
We test our formula in two lattice models: the free-fermion Qi-Wu-Zhang (QWZ) model~\cite{QWZ2006} and the Kitaev honeycomb (KH) model~\cite{KITAEV2006}. For suitable parameters, each realizes a two-dimensional chiral topological phase, with chiral central charge $\ccc=1$ and $1/2$, respectively.
Fig.~\ref{fig:num}(a) shows $-\arg Z_{g,h}/(\ccc\pi)$ for 14 representative spherical pairs $(g,h)=(g_A,g_B)$, five in $S_5$ and nine in $S_6$, whose explicit forms are given in the Supplemental Material~\cite{SM}. In every case, the results from both models agree with the prediction of Eq.~\eqref{eq:phase_Z} within a relative error of $2\%$.

To probe the approach to the thermodynamic limit, we repeat the calculation for the $S_6$ example of Eq.~\eqref{eq:exampleS6} on $2L^2$ lattice sites with $L=8,10,12,14$. As shown in Fig.~\ref{fig:num}(b), the results for both models converge toward the predicted value $1/9$ with increasing $L$, and the relative error decreases by more than an order of magnitude over this range [Fig.~\ref{fig:num}(c)].

\textbf{Discussion.---}
In this Letter, we propose a universal real-space twist-operator correlator whose phase, for spherical pairs of replica permutations, is fixed by twist-defect spins and probes the chiral central charge of a two-dimensional gapped state. Our results also raise several theoretical and numerical questions that deserve further investigation.

Fermionic signs require special care, so our path-integral interpretation of $Z_{g,h}$ and field-theoretic derivation are fully controlled only for bosonic systems. An independent fermionic derivation, perhaps adapting the free-fermion approach to the R\'{e}nyi modular commutator~\cite{gass2025renyilike,Fan2023generalized,ZangInProgress}, would therefore be valuable.

The range of validity of our formula deserves further investigation. For example, classical counterexamples~\cite{Gass2024spurious} suggest that spurious contributions could contaminate tripartite probes of the chiral central charge. Numerical tests in more general interacting lattice models are also needed. For a large number of replicas, however, variational Monte Carlo estimates of this class of real-space observables converge very slowly. Efficient numerical methods applicable to practical many-body wave functions therefore remain to be developed.

Our formula takes a compact form in terms of the topological spins of twist defects. Its structure suggests a possible connection to defect braiding, although the precise relation remains unclear. It would also be interesting to relate our construction to other lattice studies involving topological spin, including Sopenko's proposal for determining the chiral central charge~\cite{sopenko2024_1,sopenko2025_2} and studies of the statistics of higher-dimensional objects~\cite{Kobayashi2026GenStat}.

\textbf{Acknowledgments.---}
We thank Chao Xu and Hanyu Xue for useful comments, and Shenghan Jiang for useful discussions and feedback. The numerical code in this work was developed with reference to the open-source implementation by Y. Sheffer~\footnote{\url{https://github.com/ysheffer/multi-entropy-numerics}}.
This work is supported by MOST NO. 2022YFA1403902, NSFC Grant No.12574179. 
\bibliography{bibpaper}

\clearpage
\setcounter{equation}{0}
\renewcommand{\theequation}{S\arabic{equation}}
\renewcommand{\theHequation}{S\arabic{equation}}

\setcounter{table}{0}
\renewcommand{\thetable}{S\arabic{table}}
\renewcommand{\theHtable}{\thetable}

\setcounter{figure}{0}
\renewcommand{\thefigure}{S\arabic{figure}}
\renewcommand{\theHfigure}{\thefigure}

\setcounter{section}{0}
\setcounter{secnumdepth}{3}


\title{Supplemental Material for ``Chiral Central Charge from Real-Space Twist Operator Correlator''}

\maketitle

\onecolumngrid
\tableofcontents

In this Supplemental Material, we present a constructive proof of the universal phase formula stated in the main text (Sec.~\ref{sec:proof}), an analysis of the spherical condition (Sec.~\ref{sec:spherical}), and details of the numerical tests (Sec.~\ref{sec:numerics}).

\section{Constructive proof of the universal phase formula}
\label{sec:proof}
We prove the phase formula under the spherical condition $\ell(g)+\ell(h)=\ell(gh)$. We begin with the case $\ell(gh)=N-1$, for which $gh$ is an $N$-cycle. Each surface $\Sigma_v$ then contains a unique ramification-$N$ puncture, whereas every other puncture has ramification index $k<N$. The proof proceeds in three steps.

\emph{(i) Single POP.---}
Suppose that each of $g$, $h$, and $gh$ contains exactly one nontrivial cycle. Every $\Sigma_v$ is then a three-punctured sphere, i.e., a pair-of-pants surface (POP), for which the phase contribution of each connecting cylinder is known. We refer to this configuration as a single POP.

\emph{(ii) Local POP.---}
For a general spherical pair $(g,h)$ with $gh$ an $N$-cycle, a surface $\Sigma_v$ may have more than three punctures, but it still has a unique ramification-$N$ puncture located at infinity. We choose a POP decomposition of $\Sigma_v$. When gluing a puncture between two surfaces, the local POP containing that puncture serves as the reference for measuring the twist angle. 
Because the ramification-$N$ puncture is unique, it fixes the orientation, so the phase contribution of every puncture follows with a definite sign from the result of step (i).

\emph{(iii) Bulk contribution.---}
We finally show that the bulk contributions of the four surfaces $\Sigma_v$ combine into a positive real factor, which therefore does not contribute to the phase.

These three steps establish the formula for $\ell(gh)=N-1$. When $n_{\mathrm{cyc}}(gh)>1$, applying the same construction to each cycle of $gh$ yields the general spherical case.

\subsection{Step 1: Single POP}
\begin{figure}[t]
    \centering
   \includegraphics{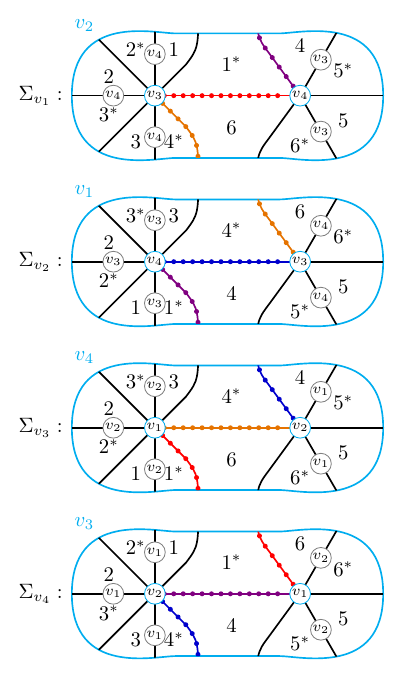}
\caption{The four surfaces $\Sigma_v$ for $g=(1234)$, $h=(456)$, and $gh=(123456)$, corresponding to $m=3$ and $n=2$. The cycle types of the relative permutations are $[g_{AB}]=[g_{CD}]=[6]$, $[g_{AC}]=[g_{BD}]=[3,1,1,1]$, and $[g_{BC}]=[g_{AD}]=[4,1,1]$; their entries give the ramification indices of the punctures. The colored dotted curves mark the global seams used to determine the twist angles.}
    \label{fig:sm-single-pop}
\end{figure}
We first consider the case in which each of $g$, $h$, and $gh$ contains exactly one nontrivial cycle. Without loss of generality, let
\begin{equation}
  \begin{aligned}
    &g=(1,2,\dots,m+1),\\
    &h=(m+1,m+2,\dots,m+n+1),\\
    &gh=(1,2,3,\dots,m+n+1).
  \end{aligned}
\end{equation}
For these permutations, each of the four surfaces $\Sigma_v$ is homeomorphic to a POP whose punctures have ramification indices $m+1$, $n+1$, and $m+n+1$. On each $\Sigma_v$, we choose a seam connecting the ramification-$(m+1)$ and ramification-$(n+1)$ punctures and continue this choice consistently across all four surfaces. These global seams follow the boundaries between the patch pairs labeled by $m+n+1|1^*$, $m+1|1^*$, $m+n+1|(m+1)^*$, and $m+1|(m+1)^*$. Fig.~\ref{fig:sm-single-pop} illustrates this construction.

We determine the twist angles by homotoping each global seam to the horizontal orthogeodesic that defines the zero-twist reference. Consider first the gluing $\Sigma_{v_1}\leftrightarrow\Sigma_{v_3}$. The red seam is already horizontal on $\Sigma_{v_1}$. On $\Sigma_{v_3}$, bringing the corresponding seam from $6|1^*$ to $2|2^*$ into the horizontal position crosses three patches and yields a twist angle of $3\pi/4$ with respect to the puncture $v_1$. Similarly, for the gluing $\Sigma_{v_1}\leftrightarrow\Sigma_{v_4}$, moving the red seam on $\Sigma_{v_4}$ from $6|1^*$ to $5|6^*$ yields a twist angle of $2\pi/3$.
Finally, the gluing $\Sigma_{v_1}\leftrightarrow\Sigma_{v_2}$ requires moving the orange seam from $6|4^*$ to $2|3^*$ on $\Sigma_{v_1}$ and from $6|4^*$ to $5|6^*$ on $\Sigma_{v_2}$. These motions cross $3+2=5$ patches in total and yield a twist angle of $-5\pi/6$; the minus sign arises because the outer boundary is used as the reference, which reverses the orientation of the sweep. The remaining three twist angles follow analogously.

In general, the twist angles are
\begin{equation}
    \begin{aligned}
       & \tau_{v_1 v_2} = \tau_{v_3 v_4} = -\pi \frac{m+n}{m+n+1}, \quad\quad\tau_{v_1 v_3} = \tau_{v_2 v_4} = \pi \frac{m}{m+1}, \quad\quad\tau_{v_1 v_4} = \tau_{v_2 v_3} = \pi \frac{n}{n+1}.
    \end{aligned}
\end{equation}
Consequently, the phase agrees with the general formula:
\begin{equation}
  \begin{aligned}
 \frac{Z_{g,h}}{\abs{Z_{g,h}}}&= \exp\left\{-\ii\frac{2\pi}{24}\ccc\left(\frac{m}{m+1}+\frac{n}{n+1}-\frac{m+n}{m+n+1}\right)\right\}\\
 &=\exp\left\{-\ii\frac{2\pi}{24}\ccc\left(m+1-\frac{1}{m+1}+n+1-\frac{1}{n+1}-m-n-1+\frac{1}{m+n+1}\right)\right\}.
  \end{aligned}
\end{equation}
Step (i) reproduces the result previously obtained for the R\'{e}nyi modular commutator~\cite{gass2025renyilike,zang2026}.

\subsection{Step 2: Local POP}
\begin{figure*}[t]
  \centering
  \includegraphics{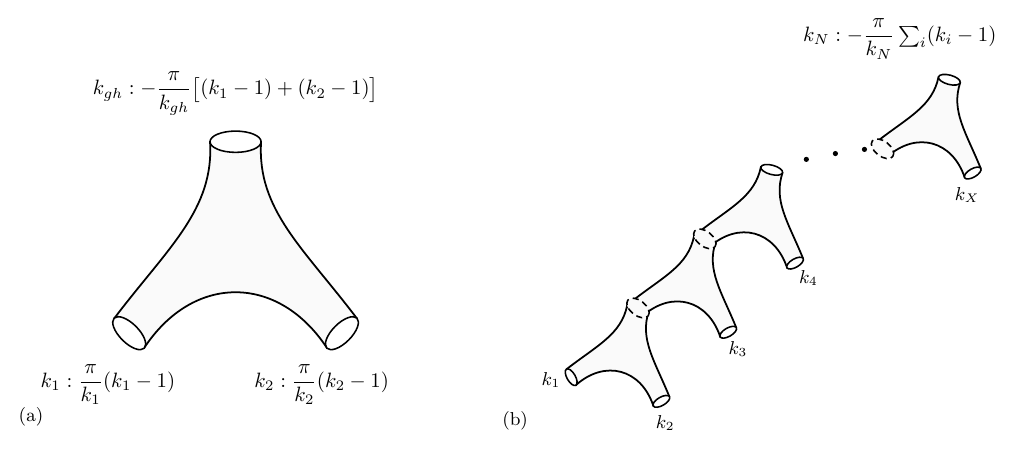}
  \caption{Pair-of-pants construction used to determine the twist angles. (a) A single POP with lower-ramification punctures $k_1$ and $k_2$. (b) A generic punctured sphere with several lower-ramification punctures $k_1,k_2,\dots,k_X$ and the highest-ramification puncture $k_N$. The black dashed curves indicate a POP decomposition.}
  \label{fig:sm-highest-ramification-puncture}
\end{figure*}
We next allow $g$ and $h$ to have arbitrary cycle decompositions while retaining the assumption that $gh$ is an $N$-cycle. Each $\Sigma_v$ is therefore a generic punctured sphere, and the CFT partition function on the total surface $\Sigma$ can be written as 
\begin{equation}
  \calZ(\Sigma)=\ee^{-\ccc\frac{2\pi}{24}\sum_{IJ}\tau_{IJ}} \prod_v \calZ(\Sigma_v;\{\ket{0}_p\})
\end{equation}
where $IJ\in\{AB,BC,AC,AD,BD,CD\}$ labels the interfaces, and $\calZ(\Sigma_v;\{\ket{0}_p\})$ denotes the CFT partition function on the punctured sphere $\Sigma_v$ with vacuum boundary conditions $\ket{0}_p$ at all punctures. The phase factor $\ee^{-\ccc\frac{2\pi}{24}\tau_{IJ}}$ arises from gluing the punctures that lie on the interface $IJ$. In this step, we determine the twist angles $\tau_{IJ}$.

\begin{figure*}[t]
  \centering
    \includegraphics{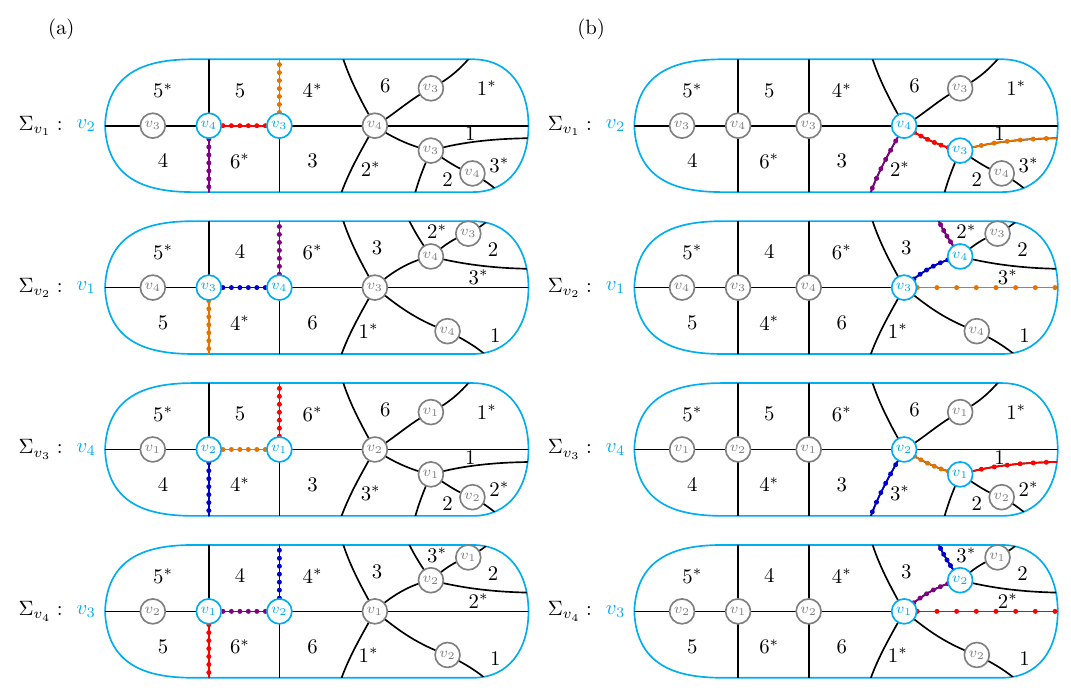}
  \caption{Local-POP constructions for the four surfaces $\Sigma_{v_1}$--$\Sigma_{v_4}$ with
        $g_A=(23)(46)$, $g_B=(136)(45)$, and $g_C=(123456)$.
        The cycle types of the relative permutations are
        $[g_{AB}]=[g_{CD}]=[6]$,
        $[g_{AC}]=[g_{BD}]=[3,2,1]$, and
        $[g_{BC}]=[g_{AD}]=[2,2,1,1]$.
        Panels (a) and (b) show two complementary choices of local POPs; in each panel, the three cyan vertices mark the punctures contained in the local POP used as the reference. Applying step (i) to these local POPs determines the twist angles.}
  \label{fig:local-POP_}
\end{figure*}

Each $\Sigma_v$ has a unique ramification-$N$ puncture, hereafter called the highest-ramification puncture, which lies at infinity; all other punctures have ramification index $k<N$ and lie at finite positions.
We choose a POP decomposition of each $\Sigma_v$, as shown in Fig.~\ref{fig:sm-highest-ramification-puncture}(b). When gluing a puncture between two surfaces, the local POP containing that puncture serves as the reference. Every local POP inherits this property: its punctures with $k<N$ lie at finite positions, while its highest-ramification puncture can be homotoped to infinity. Applying the result of step (i) to the ``old'' punctures (as opposed to the ``new'' punctures generated by the POP decomposition) then gives the twist angle associated with a ramification-$k$ puncture:
\begin{equation}
  \tau_k =\left\{
  \begin{aligned}
     &\pi\frac{k-1}{k},\qquad k<N,\\
     &-\pi\frac{N-1}{N},\qquad k=N.
  \end{aligned}\right.
\end{equation}
An example of this procedure is shown in Fig.~\ref{fig:local-POP_}.

Summing the resulting phase factors yields the phase formula for a spherical pair with $gh$ an $N$-cycle:
\begin{equation}
  \begin{aligned}
    \arg Z_{g,h}&= -\frac{2\pi}{24}\ccc \left[\left(\sum_{c\in g,h}1-\frac{1}{\abs{c}}\right)-\left(1-\frac{1}{N}\right) \right] \\
    &=-\frac{2\pi}{24}\ccc \left[\left(\sum_{c\in g,h}1-\frac{1}{\abs{c}}\right)+\left(\sum_{c\in g,h}\abs{c}-1\right)-(N-1)-\left(1-\frac{1}{N}\right) \right] \\
     &=-\frac{2\pi}{24}\ccc \left[\left(\sum_{c\in g,h}\abs{c}-\frac{1}{\abs{c}}\right)-\left(N-\frac{1}{N}\right) \right] \\
  \end{aligned}
\end{equation}
where the second equality uses the spherical condition,
\begin{equation}
  \sum_{c\in g,h} (\abs{c}-1)=N-1.
\end{equation}
The final expression is precisely the phase of $(\theta_g\theta_h/\theta_{gh})^{\ccc}$ for an $N$-cycle $gh$.

\subsection{Step 3: Bulk contribution}
Finally, we show that the bulks of the four surfaces $\Sigma_v$ do not contribute to the phase. Let $\calV$ denote the CFT partition function on a punctured sphere with vacuum boundary conditions; keeping track of the orientation, it depends on three permutations:  
\begin{equation}
  \begin{aligned}
    &\textrm{ABC:}\quad\calZ(\Sigma_{v_1};\{\ket{0}_p\})=\calV(g,h,gh)\\
     &\textrm{ABD:}\quad\calZ(\Sigma_{v_2};\{\ket{0}_p\})=\calV(h,g,e)\\
      &\textrm{ACD:}\quad\calZ(\Sigma_{v_3};\{\ket{0}_p\})=\calV(g,gh,e)\\
       &\textrm{BCD:}\quad\calZ(\Sigma_{v_4};\{\ket{0}_p\})=\calV(h,e,gh)
  \end{aligned}
\end{equation}
A global relabeling of the replicas leaves the partition function invariant, and inverting all three permutations complex-conjugates it:
\begin{equation}
  \begin{aligned}
   & \calV(g,h,gh) = \calV(g^{-1}gh^{-1},g^{-1}hh^{-1},g^{-1}ghh^{-1}) = \calV(h^{-1},g^{-1},e) =\overline{\calV(h,g,e)} \\
   & \calV(g,gh,e) = \calV((gh)^{-1}g,(gh)^{-1}gh,(gh)^{-1}e) = \calV(h^{-1},e,(gh)^{-1})=\overline{\calV(h,e,gh)} \\
  \end{aligned}
\end{equation}
Consequently, the four bulk partition functions multiply to $\abs{\calV(h,g,e)}^{2}\,\abs{\calV(h,e,gh)}^{2}$, which is real and positive and hence contributes no phase. Together with steps (i) and (ii), this completes the proof of the phase formula for $\ell(gh)=N-1$.

\section{The spherical condition}
\label{sec:spherical}
In this section, we analyze the spherical condition in detail. We first show that a spherical pair factorizes, so that it suffices to consider the case $n_{\mathrm{cyc}}(gh)=1$. We then show that the three-manifold specified by such a pair is in fact the three-sphere.
\subsection{Factorization of a spherical pair}
The transposition length of a permutation $g\in S_N$ can be written as
\begin{equation}
  \ell(g)\equiv N-n_{\mathrm{cyc}}(g) \equiv \sum_{c\subset g} (\abs{c}-1),\quad g\in S_N.
\end{equation}
We now characterize the pairs $(g,h)\in S_N\times S_N$ that satisfy the spherical condition
\begin{equation}
  \ell(g)+\ell(h)=\ell(gh).
\end{equation}
Choose minimal transposition decompositions of $g$ and $h$,
\begin{equation}
  \begin{aligned}
    & g= t_1 t_2\cdots t_{\ell(g)},\\
    & h=s_1 s_2 \cdots s_{\ell(h)},
  \end{aligned}
\end{equation}
where all $t_i$ and $s_j$ are transpositions. The spherical condition implies that the concatenated word
\begin{equation}
  gh=  t_1 t_2\cdots t_{\ell(g)}s_1 s_2 \cdots s_{\ell(h)},
\end{equation}
is also a minimal transposition decomposition of $gh$. We associate with this decomposition a graph whose vertices and edges are
\begin{equation}
  \begin{aligned}
    &N~\mathrm{vertices:}~1,2,3,\dots,N, \\
    &N-n_{\mathrm{cyc}}(gh)~\mathrm{edges:}~ t_i,s_j.
  \end{aligned}
\end{equation}
Let $Y$ be the number of connected components of this graph.

\emph{Claim.---} The connected components are in one-to-one correspondence with the cycles of $gh$, i.e.,
\begin{equation}
  \boxed{Y=n_{\mathrm{cyc}}(gh).}
\end{equation}

\emph{Proof.---} Starting from $N$ isolated vertices, adding one edge can reduce the number of connected components by at most one. The graph therefore has at least $N-(N-n_{\mathrm{cyc}}(gh))$ connected components, and hence $Y\ge n_{\mathrm{cyc}}(gh)$.

Conversely, every transposition $t_i$ or $s_j$ acts within a connected component, so the action of $gh$ preserves each component. Every cycle of $gh$ is therefore contained in a connected component, and each component contains at least one such cycle. It follows that $Y\le n_{\mathrm{cyc}}(gh)$. Combining the two inequalities completes the proof. $\square$

Write the cycle decomposition of $gh$ as
\begin{equation}
  gh =c_1 c_2\cdots c_{n_{\mathrm{cyc}}(gh)}.
\end{equation}
The supports $\mathrm{supp}(c_1),\mathrm{supp}(c_2),\dots,\mathrm{supp}(c_{n_{\mathrm{cyc}}(gh)})$ partition the vertices into these connected components. Accordingly, $g$ and $h$ factorize as
\begin{equation}
  \begin{aligned}
    & g= g_1 g_2\dots g_{n_{\mathrm{cyc}}(gh)}, \\
    & h= h_1 h_2\dots h_{n_{\mathrm{cyc}}(gh)},
  \end{aligned}
\end{equation}
where $g_k$ and $h_k$ are supported on $\mathrm{supp}(c_k)$ and need not themselves be cycles or transpositions. The identity $g\cdot h=gh$ requires
\begin{equation}
  g_k h_k = c_k,\quad k=1,2,\dots,n_{\mathrm{cyc}}(gh),
\end{equation}
and therefore
\begin{equation}
  \ell(g_k)+\ell(h_k) = \ell(c_k),\quad k=1,2,\dots,n_{\mathrm{cyc}}(gh),\quad g_k,h_k,c_k\in S_{\abs{c_k}}.
\end{equation}
Thus, the problem factorizes over the cycles of $gh$, and it is sufficient to consider the case $n_{\mathrm{cyc}}(gh)=1$.
\subsection{Geometric interpretation}

We next give a geometric interpretation of the spherical condition and show
that, in the genuinely $N$-sheeted sector,
\begin{equation}
  \ell(g)+\ell(h)=\ell(gh),
  \qquad
  n_{\mathrm{cyc}}(gh)=1,
  \label{eq:spherical-condition}
\end{equation}
the corresponding replica path integral is defined on the three-sphere
$S^3$.

Recall that
\begin{equation}
  Z_{g,h}
  =
  \mathrm{Tr}_{ABC}\left[
  \rho_{ABC}^{\otimes N}
  \TT_A(g)\TT_B(h)\TT_C(g)\TT_C(h)
  \right],
  \qquad
  \rho_{ABC}
  =
  \mathrm{Tr}_D\ket{\psi}\bra{\psi}.
\end{equation}
We place the system on the spatial manifold $S^2$. The Euclidean path
integral preparing the ground state $\ket{\psi}$ fills $S^2$ with a
three-ball $D^3$, whose boundary is
\begin{equation}
  \partial D^3=S^2.
\end{equation}
We divide the boundary into the upper hemisphere $ABC$ and the lower
hemisphere $D$. The density matrix $\ket{\psi}\bra{\psi}$ is represented
by a ket three-ball and an orientation-reversed bra three-ball. Taking
the partial trace over $D$ glues the two balls along their $D$ regions.
The resulting manifold representing $\rho_{ABC}$ is therefore
topologically again a three-ball, whose boundary consists of the two
remaining copies,
\begin{equation}
  \partial D^3
  =
  ABC\cup A^*B^*C^*.
\end{equation}
As the simplest example, $\mathrm{Tr}_{ABC}\rho_{ABC}$ further identifies
$ABC$ with $A^*B^*C^*$ and hence gives
\begin{equation}
  D^3\cup_{S^2}\overline{D^3}
  \simeq S^3.
\end{equation}

Following Ref.~\cite{sheffer2026single}, we now prove that the path-integral representation of $Z_{g,h}$ is defined on $S^3$ using methods from combinatorial topology. The replica gluing associated with $Z_{g,h}$ can be encoded by a bipartite $4$-colored graph $\GG$.
The graph contains $2N$ vertices,
\begin{equation}
  V(\GG)
  =
  \{1,1^*,2,2^*,\ldots,N,N^*\},
\end{equation}
where $r$ and $r^*$ correspond to the ket and bra copies, respectively.
The four edge colors are labeled by $R=A,B,C,D$, and an $R$-colored
edge connects two vertices:
\begin{equation}
  r\longleftrightarrow g_R(r)^*.
\end{equation}
For the present construction,
\begin{equation}
  g_A=g,\qquad
  g_B=h,\qquad
  g_C=gh,\qquad
  g_D=e.
  \label{eq:four-permutations}
\end{equation}
Here the $D$-colored edges arise from the trace over $D$ within each
replica. Below we show that the geometry encoded by this graph is (i) a closed three-manifold and (ii) in fact the three-sphere $S^3$.

For each color $R$, let $\GG_R$ denote the $3$-colored residue obtained
by removing all $R$-colored edges from $\GG$. A $4$-colored graph
encodes a closed PL $3$-manifold if every connected component of every
$\GG_R$ triangulates $S^2$. In this case, $\GG$ is called a
$3$-gem.

For later use, we also define $\GG_{IJ}$ as the $2$-colored residue
obtained by removing all edges of colors $I$ and $J$. Since every vertex
of $\GG_{IJ}$ has degree two, each connected component of $\GG_{IJ}$ is
a closed alternating two-colored loop. We denote the number of such
loops by $\abs{\GG_{IJ}}$.

\emph{Lemma~\cite{ryan2016colored}.---} For four distinct colors $I,J,K,L$, the conditions
\begin{align}
  \abs{\GG_{IJ}}
  &=
  \abs{\GG_{KL}},
  \label{eq:gem-condition-1}
  \\
  \abs{\GG_{AB}}
  +
  \abs{\GG_{BC}}
  +
  \abs{\GG_{AC}}
  &=
  N+2,
  \label{eq:gem-condition-2}
\end{align}
are sufficient to guarantee that the present $4$-colored graph is a
$3$-gem.

The two-colored residues can be evaluated directly from the permutation
data. Suppose that $\GG_{IJ}$ retains colors $K$ and $L$. Starting from
an unstarred vertex $r$ and following successively a $K$-colored edge
and an $L$-colored edge gives
\begin{equation}
  r
  \longrightarrow
  g_K(r)^*
  \longrightarrow
  g_L^{-1}g_K(r).
\end{equation}
Repeatedly traversing the two-colored loop therefore traces a cycle of
the permutation $g_L^{-1}g_K$. Consequently,
\begin{equation}
  \abs{\GG_{IJ}}
  =
  n_{\mathrm{cyc}}\!\left(g_L^{-1}g_K\right)
  =
  n_{\mathrm{cyc}}\!\left(g_K^{-1}g_L\right).
  \label{eq:two-color-cycle}
\end{equation}
Using Eq.~\eqref{eq:four-permutations}, we obtain the numbers of connected components summarized in Table~\ref{tab:two-colored-residues}.
\begin{table}[t]
\centering
\begin{tabular}{ccc}
\toprule
Residue & Remaining colors & $\abs{\GG_{IJ}}$ \\
\midrule
$\GG_{AB}$ & $C,D$ & $n_{\mathrm{cyc}}(gh)$ \\
$\GG_{CD}$ & $A,B$ & $n_{\mathrm{cyc}}(g^{-1}h)$ \\
$\GG_{AC}$ & $B,D$ & $n_{\mathrm{cyc}}(h)$ \\
$\GG_{BD}$ & $A,C$ & $n_{\mathrm{cyc}}(h)$ \\
$\GG_{AD}$ & $B,C$ & $n_{\mathrm{cyc}}(g)$ \\
$\GG_{BC}$ & $A,D$ & $n_{\mathrm{cyc}}(g)$ \\
\bottomrule
\end{tabular}
\caption{The numbers of connected components of the two-colored
residues for
$g_A=g$, $g_B=h$, $g_C=gh$, and $g_D=e$.}
\label{tab:two-colored-residues}
\end{table}

We now impose Eq.~\eqref{eq:spherical-condition}. Since
$n_{\mathrm{cyc}}(gh)=1$, the permutation $gh$ is an $N$-cycle and
hence
\begin{equation}
  \ell(gh)=N-1.
\end{equation}
The length-additivity condition therefore gives
\begin{equation}
  \ell(g)+\ell(h)=N-1,
\end{equation}
or equivalently,
\begin{equation}
  n_{\mathrm{cyc}}(g)+n_{\mathrm{cyc}}(h)=N+1.
  \label{eq:cycle-sum}
\end{equation}

We also need the cycle structure of $g^{-1}h$. Since $gh$ is an
$N$-cycle, the subgroup generated by $g$ and $h$ acts transitively on
$\{1,\ldots,N\}$. For a transitive pair of permutations, the standard
Euler-characteristic bound gives
\begin{equation}
  n_{\mathrm{cyc}}(g)
  +
  n_{\mathrm{cyc}}(h)
  +
  n_{\mathrm{cyc}}(g^{-1}h)
  \leq N+2.
\end{equation}
Using Eq.~\eqref{eq:cycle-sum}, we find
\begin{equation}
  n_{\mathrm{cyc}}(g^{-1}h)\leq 1.
\end{equation}
Since every permutation has at least one cycle,
\begin{equation}
  n_{\mathrm{cyc}}(g^{-1}h)=1.
  \label{eq:g-inv-h-cycle}
\end{equation}

Equations~\eqref{eq:cycle-sum} and \eqref{eq:g-inv-h-cycle} immediately
imply
\begin{align}
  \abs{\GG_{AB}}
  &=
  \abs{\GG_{CD}}
  =
  1,
  \\
  \abs{\GG_{AC}}
  &=
  \abs{\GG_{BD}}
  =
  n_{\mathrm{cyc}}(h),
  \\
  \abs{\GG_{AD}}
  &=
  \abs{\GG_{BC}}
  =
  n_{\mathrm{cyc}}(g),
\end{align}
which proves Eq.~\eqref{eq:gem-condition-1}. Moreover,
\begin{align}
  \abs{\GG_{AB}}
  +
  \abs{\GG_{BC}}
  +
  \abs{\GG_{AC}}
  &=
  1+n_{\mathrm{cyc}}(g)+n_{\mathrm{cyc}}(h)
  \nonumber\\
  &=
  N+2,
\end{align}
which proves Eq.~\eqref{eq:gem-condition-2}. Hence $\GG$ is a
$3$-gem and therefore encodes a closed orientable $3$-manifold, which
we denote by $M_{g,h}$.

It remains to identify the topology of $M_{g,h}$. For this purpose, we
use the jacket construction of Ref.~\cite{ryan2016colored}. A jacket
associated with a $4$-colored graph is a closed two-dimensional
orientable surface that contains all vertices and all edges of $\GG$,
while its faces are selected from the two-colored loops according to a
cyclic ordering of the four colors.

We choose the cyclic ordering
\begin{equation}
  \tau=(A,C,B,D).
\end{equation}
The corresponding jacket $J_{ACBD}$ contains the four types of
two-colored faces
\begin{equation}
  AC,\qquad
  CB,\qquad
  BD,\qquad
  DA.
\end{equation}
The numbers of these faces are, respectively,
\begin{equation}
  n_{\mathrm{cyc}}(h),\qquad
  n_{\mathrm{cyc}}(g),\qquad
  n_{\mathrm{cyc}}(h),\qquad
  n_{\mathrm{cyc}}(g).
\end{equation}
Therefore the total number of jacket faces is
\begin{align}
  F_J
  &=
  2n_{\mathrm{cyc}}(g)
  +
  2n_{\mathrm{cyc}}(h)
  \nonumber\\
  &=
  2N+2,
  \label{eq:jacket-face-number}
\end{align}
where Eq.~\eqref{eq:cycle-sum} has been used.

The jacket contains the same $2N$ vertices as $\GG$. Since $\GG$ is
$4$-regular, it has $4N$ edges. Hence
\begin{equation}
  V_J=2N,
  \qquad
  E_J=4N.
\end{equation}
Its Euler characteristic is consequently
\begin{align}
  \chi(J_{ACBD})
  &=
  V_J-E_J+F_J
  \nonumber\\
  &=
  2N-4N+(2N+2)
  \nonumber\\
  &=
  2.
\end{align}
The graph $\GG$ is connected because the $CD$ two-colored residue
consists of a single loop,
\begin{equation}
  \abs{\GG_{AB}}
  =
  n_{\mathrm{cyc}}(gh)
  =
  1,
\end{equation}
and the jacket has the same vertices and edges as $\GG$. Therefore
$J_{ACBD}$ is a connected orientable closed surface. Using
\begin{equation}
  \chi(J)=2-2g_J,
\end{equation}
we obtain
\begin{equation}
  g_J=0,
\end{equation}
and hence
\begin{equation}
  J_{ACBD}\simeq S^2.
\end{equation}

Thus $\GG$ possesses a spherical jacket. For a $4$-colored graph
encoding a closed $3$-manifold, a spherical jacket gives a genus-zero
Heegaard splitting. Equivalently, the conclusion follows from the following lemma.

\emph{Lemma~\cite{ryan2016colored}.---} A $3$-gem possessing a spherical jacket represents the three-sphere.

Applying the lemma to $\GG$ gives
\begin{equation}
  M_{g,h}\simeq S^3.
\end{equation}

We thus conclude that, in the sector $n_{\mathrm{cyc}}(gh)=1$, the
length-additivity condition
\begin{equation}
  \ell(g)+\ell(h)=\ell(gh),
\end{equation}
has a direct geometric interpretation: it guarantees that the replica
gluing defining $Z_{g,h}$ produces the three-sphere $S^3$.
\section{Numerical details}
\label{sec:numerics}
This section provides the details of the numerical tests summarized in the main text. Classifying the spherical pairs with $[gh]=[N]$ by their cycle types $([g],[h],[gh])$, modulo the exchange $g\leftrightarrow h$, yields five classes in $S_5$ (Table~\ref{tab:S5-spherical-lift-breaking}) and nine classes in $S_6$ (Table~\ref{tab:S6-spherical-lift-breaking}). For each class, the representative $(g,h,gh)\equiv(g_A,g_B,g_C)$ listed in the tables is the one used in the numerics.

\begin{table}[t]

\centering

\begin{tabular}{cccccc}

\toprule

Case & $[g]$ & $[h]$ & $[gh]$ & $\arg Z/(\ccc\pi)$ & Example $(g,h,gh)$ \\

\midrule

$\mathrm{E}_1$ & $[2]$   & $[4]$   & $[5]$ & $3/80$
& $((45),(1234),(12354))$ \\

$\mathrm{E}_2$ & $[2]$   & $[3,2]$ & $[5]$ & $13/180$
& $((45),(124)(35),(12534))$ \\

$\mathrm{E}_3$ & $[3]$   & $[3]$   & $[5]$ & $2/45$
& $((345),(123),(12453))$ \\

$\mathrm{E}_4$ & $[3]$   & $[2,2]$ & $[5]$ & $13/180$
& $((345),(13)(24),(14253))$ \\

$\mathrm{E}_5$ & $[2,2]$ & $[2,2]$ & $[5]$ & $1/10$
& $((23)(45),(12)(34),(13542))$ \\

\bottomrule

\end{tabular}

\caption{The five classes of spherical pairs in $S_5$ with $[gh]=[5]$, up to the exchange $g\leftrightarrow h$ (cases $E_1$--$E_5$ in Fig.~5 of the main text). Each representative satisfies $\ell(g)+\ell(h)=\ell(gh)=4$.
}

\label{tab:S5-spherical-lift-breaking}
\end{table}
\begin{table}[t]

\centering

\begin{tabular}{cccccc}

\toprule

Case & $[g]$ & $[h]$ & $[gh]$ & $\arg Z/(\ccc\pi)$ & Example $(g,h,gh)$ \\

\midrule

$\mathrm{E}_6$ & $[2]$ & $[5]$ & $[6]$ & $7/180$
& $((56),(12345),(123465))$ \\

$\mathrm{E}_7$ & $[2]$ & $[4,2]$ & $[6]$ & $11/144$
& $((56),(1235)(46),(123645))$ \\

$\mathrm{E}_8$ & $[2]$ & $[3,3]$ & $[6]$ & $1/12$
& $((56),(125)(346),(126345))$ \\

$\mathrm{E}_9$ & $[3]$ & $[4]$ & $[6]$ & $7/144$
& $((456),(1234),(123564))$ \\

$\mathrm{E}_{10}$ & $[3]$ & $[3,2]$ & $[6]$ & $1/12$
& $((456),(124)(35),(125364))$ \\

$\mathrm{E}_{11}$ & $[3]$ & $[2,2,2]$ & $[6]$ & $1/9$
& $((456),(14)(25)(36),(152634))$ \\

$\mathrm{E}_{12}$ & $[2,2]$ & $[4]$ & $[6]$ & $11/144$
& $((34)(56),(1235),(124365))$ \\

$\mathrm{E}_{13}$ & $[2,2]$ & $[3,2]$ & $[6]$ & $1/9$
& $((34)(56),(123)(45),(124653))$ \\

$\mathrm{E}_{14}$ & $[2,2]$ & $[2,2,2]$ & $[6]$ & $5/36$
& $((34)(56),(13)(25)(46),(145263))$ \\

\bottomrule

\end{tabular}

\caption{The nine classes of spherical pairs in $S_6$ with $[gh]=[6]$, up to the exchange $g\leftrightarrow h$ (cases $E_6$--$E_{14}$ in Fig.~5 of the main text). Each representative satisfies $\ell(g)+\ell(h)=\ell(gh)=5$.
}

\label{tab:S6-spherical-lift-breaking}

\end{table}
\subsection{Free fermion}
We work with complex fermionic modes satisfying $\{c_i^\dagger,c_j\}=\delta_{ij}$. We first derive the explicit form of the twist operator. For $g\in S_N$, write $\TT(g)=\ee^{c_r^\dagger (M_g)_{rr^\prime}c_{r^\prime}}$, where $M_g$ is an $N\times N$ matrix in replica space and repeated replica indices are summed over. Let $g=c_1c_2\cdots c_X$ be the disjoint-cycle decomposition. For a cycle $c=(r_1r_2\cdots r_Y)$ of $g$, the corresponding submatrix of $\ee^{M_g}\equiv R_g$ is
\begin{equation}
  (\ee^{M_g})_{r_1 r_2\cdots r_Y}
  \equiv (R_g)_{r_1 r_2\cdots r_Y}
  =
  \begin{pmatrix}
    0 & 0 & \cdots & 0 & 1 \\
    -1 & 0 & \cdots & 0 & 0 \\
    0 & -1 & \ddots & 0 & 0 \\
    \vdots & & \ddots & \ddots & \vdots \\
    0 & 0 & \cdots & -1 & 0
  \end{pmatrix}.
\end{equation}
This can be checked using the relation
\begin{equation}
\TT(g) c_r \TT(g)^{-1} = \sum_{r^\prime} (\ee^{-M_g})_{rr^\prime} c_{r^\prime}=\sum_{r^\prime} (R_g^{-1})_{rr^\prime} c_{r^\prime}.
\end{equation}
As in the main text, we choose $r_1$ to be the smallest element of the cycle $c$, which fixes the sign convention of the twist operator unambiguously.

For the lattice site $\vec{x}$ and replica index $r$, the twist operator in region $A$ is
\begin{equation}
  \TT_A(g)= \ee^{\sum_{\vec{x}\in A}c_{\vec{x},r}^\dagger (M_g)_{rr^\prime}c_{\vec{x},r^\prime} } =\ee^{F^\dagger (M_g\otimes\Pi_A) F}
\end{equation}
where $\Pi_A$ is the projector onto region $A$, i.e., $(\Pi_A)_{\vec{x}\vec{x}^\prime}=\delta_{\vec{x}\vec{x}^\prime}$ for $\vec{x}\in A$ and zero otherwise, and $F_{\vec{x},r}$ collects the fermionic modes in the tensor product of the real and replica spaces.

Zero-temperature expectation values of such fermionic Gaussian operators can be evaluated with the Klich formula~\cite{Peschel2003,Klich2003}. For the free-fermion ground state $\ket{\psi}$, define the spectral projector matrix $P$ by
\begin{equation}
  P_{\vec{x}\vec{x}^\prime} = \bra{\psi}c_{\vec{x}^\prime}^\dagger c_{\vec{x}}\ket{\psi}.
\end{equation}
A Gaussian operator then has the expectation value
\begin{equation}
  \bra{\psi} \ee^{c_{\vec{x}}^\dagger M_{\vec{x}\vec{x}^\prime}c_{\vec{x}^\prime}} \ket{\psi}=\det \left[\Id-P +P\ee^M P\right]
\end{equation}
For the tripartite observable $Z_{g,h}$ of Eq.~(1) of the main text, this gives
\begin{equation}
   Z_{g,h}=\bra{\psi^{(N)}} \TT_A(g)\TT_B(h)\TT_C(g)\TT_C(h)\ket{\psi^{(N)}}=\det\left[\Id-\Id_N\otimes P+(\Id_N\otimes P)\ee^{M_g\otimes\Pi_A}\ee^{M_h\otimes\Pi_B}\ee^{M_g\otimes\Pi_C}\ee^{M_h\otimes\Pi_C}(\Id_N\otimes P)\right] 
\end{equation}

\emph{Majorana fermion.---} For Majorana modes satisfying $\{\gamma_i,\gamma_j\}=2\delta_{ij}$, the twist operator is
\begin{equation}
\TT(g)=\ee^{\frac{1}{4}\gamma_r (M_g)_{rr^\prime}\gamma_{r^\prime}},
\end{equation}
where $M_g$ is antisymmetric, $M_g^T=-M_g$. Its action on the Majorana operators follows from the Baker--Campbell--Hausdorff expansion:
\begin{equation}
\TT(g)\gamma_r\TT(g)^{-1}=(\ee^{-M_g})_{rr^\prime}\gamma_{r^\prime}=(R_g^{-1})_{rr^\prime}\gamma_{r^\prime},
\end{equation}
with the same convention $\ee^{M_g}\equiv R_g$ as above.

The Majorana spectral projector is defined by
\begin{equation}
  2P_{\vec{x}\vec{x}^\prime} = \bra{\psi}\gamma_{\vec{x}^\prime}\gamma_{\vec{x}}\ket{\psi},
\end{equation}
and a Majorana Gaussian operator has the zero-temperature expectation value~\cite{Klich2014paired}
\begin{equation}
  \bra{\psi} \ee^{\frac{1}{4}\gamma_{\vec{x}} M_{\vec{x}\vec{x}^\prime}\gamma_{\vec{x}^\prime}} \ket{\psi}=\pm\det^{\frac{1}{2}} \left[\Id-P +P\ee^M P\right].
\end{equation}
Thus, free Majorana and complex fermion systems can be treated on the same footing, except for the sign ambiguity of the square root.

\emph{QWZ model.---}
The two-dimensional Qi-Wu-Zhang (QWZ) model~\cite{QWZ2006} is a prototypical lattice model of a Chern insulator, with Hamiltonian
\begin{equation}
    H=\sum_{\vec{k}} \begin{pmatrix}
        a_{\vec{k}}^\dagger&b_{\vec{k}}^\dagger
    \end{pmatrix}
   \vec{h}(\vec{k})\cdot \vec{\sigma}
    \begin{pmatrix}
        a_{\vec{k}} \\
        b_{\vec{k}}
    \end{pmatrix},
\end{equation}
where $\vec{h}(\vec{k}) = (\sin k_x, \sin k_y, M - \cos k_x - \cos k_y)$ and $\vec{\sigma}$ is the vector of Pauli matrices. The model realizes a Chern insulator phase for $|M| < 2$. It is a two-band model defined on an $L\times L$ square lattice ($L^2$ sites, i.e., $2L^2$ fermionic modes) with periodic boundary conditions. We take the mass parameter $M=1$, in the middle of the $\ccc=1$ Chern phase, and add weak random bond disorder on the $x$-links, drawn independently and uniformly from $[0,0.5)$, to move the ground state away from the clean limit. The disk $A\cup B\cup C$ is a $(2R+1)\times(2R+1)$ square centered on the lattice with $R=\lfloor 5L/14\rfloor$, partitioned into three regions as in Fig.~1(a) of the main text. The 14 cases in Fig.~5(a) of the main text are computed at $L=14$ ($R=5$, i.e., $121$ sites in $A\cup B\cup C$), while the size scan in Figs.~5(b) and 5(c) of the main text uses $L=8,10,12,14$.

\subsection{Kitaev honeycomb model}
The Kitaev honeycomb (KH) model is an interacting spin model that is solved by Majorana partons, $\{\sigma_x,\sigma_y,\sigma_z\}\to \{c,b^x,b^y,b^z\}$; for details of the model, see Ref.~\cite{KITAEV2006}. The tripartite observable is evaluated following Ref.~\cite{sheffer2025probing}, and we briefly review the procedure below:
\begin{enumerate}
  \item Construct the ground state of the $c$ fermions together with the dimer state of the $b^{x,y,z}$ fermions, denoted $\ket{\psi}$.
  \item Impose the gauge-invariant constraint $\ket{\psi}\to\prod_{j}\frac{1+G_{j}}{2}\ket{\psi}$ with $G_{j}=b_{j}^x b_{j}^y b_{j}^z c_{j}$.
  \item The tripartite observable is
   \begin{equation}
   \frac{ \bra{\psi^{(N)}} \prod_{j,r}\frac{1+G_{j,r}}{2}\pi_A(g_A) \pi_B(g_B) \pi_C(g_C)\ket{\psi^{(N)}}} {\bra{\psi^{(N)}} \prod_{j,r}\frac{1+G_{j,r}}{2}\ket{\psi^{(N)}}}=\frac{1}{2^N}\sum_{\{s_{I,r}=0,1\}}\bra{\psi^{(N)}} \prod_{I,r}G_{I,r}^{s_{I,r}} \pi_A(g_A) \pi_B(g_B) \pi_C(g_C) \ket{\psi^{(N)}},
   \label{eq:kh-gauge-sum}
\end{equation}
where $I=A,B,C,D$ and $r=1,2,\dots,N$.
\item For each allowed set $\{s_{I,r}\}$, evaluate the corresponding zero-temperature expectation value in the free Majorana state. The sign ambiguity of the square root is fixed by choosing the default branch in the code, which is justified by the smallness of the resulting phase.
\end{enumerate}
Here $\pi_R(g)$ denotes a choice of lift of the spin permutation $g$ to the Majorana representation in region $R$, and different choices are equivalent thanks to the gauge-invariant projection.

For the Kitaev honeycomb model we use isotropic couplings $J_x=J_y=J_z=1$ and the next-nearest-neighbor coupling $\kappa=0.2$, which places the model in the gapped non-Abelian phase with $\ccc=1/2$. The model is defined on an $L\times L$ honeycomb lattice with $2L^2$ sites and periodic boundary conditions, and the disk $A\cup B\cup C$ is a circle of radius $0.4$ times the linear lattice extent, divided into three $2\pi/3$ sectors. The 14 cases in Fig.~5(a) of the main text are computed at $L=10$.

\emph{Allowed sets $\{s_{I,r}\}$.---}
We derive the conditions selecting the nonvanishing terms in Eq.~\eqref{eq:kh-gauge-sum}, following Ref.~\cite{sheffer2025probing}, using the untwisted lift $\pi_R(g)\, b_r\,\pi_R(g)^{-1} = b_{g[r]}$; the final physical observable is independent of this choice. Consider a dimer bond crossing the boundary between two regions $I$ and $J$, whose Majoranas $b_{I,r}$, $b_{J,r}$ on replica $r$ satisfy $\ii b_{I,r}b_{J,r}\ket{\psi}=\ket{\psi}$. Focusing on these bonds, a general term can be manipulated as
\begin{equation}
    \begin{aligned}
        &\quad\quad\Tr\!\left[\prod_r G_{I,r}^{s_{I,r}}G_{J,r}^{s_{J,r}} \pi_I(g_I) \pi_J(g_J)\prod_r \frac{1+\ii b_{I,r}b_{J,r}}{2}\right] \\
        &=\Tr\!\left[\prod_r G_{I,r}^{s_{I,r}}G_{J,r}^{s_{J,r}} \pi_I(g_I) \pi_J(g_J)\left(\prod_t \ii b_{I,t}b_{J,t}\right)\left(\prod_r \frac{1+\ii b_{I,r}b_{J,r}}{2}\right)\right]\\
         &=\Tr\!\left[\prod_r G_{I,r}^{s_{I,r}}G_{J,r}^{s_{J,r}} \left(\prod_t \ii b_{I,g_I(t)}b_{J,g_J(t)}\right)\pi_I(g_I) \pi_J(g_J)\left(\prod_r \frac{1+\ii b_{I,r}b_{J,r}}{2}\right)\right]\\
         &=\Tr\!\left[\prod_r G_{I,r}^{s_{I,r}}G_{J,r}^{s_{J,r}} \left(\prod_q \ii b_{I,g_{IJ}(q)}b_{J,q}\right)\pi_I(g_I) \pi_J(g_J)\left(\prod_r \frac{1+\ii b_{I,r}b_{J,r}}{2}\right)\right],
    \end{aligned}
\end{equation}
where $g_{IJ}\equiv g_I g_J^{-1}$: we first insert $\prod_t \ii b_{I,t}b_{J,t}$, which is the identity on the dimer-projected state, then commute the twists through, and finally relabel $q=g_J(t)$. Restricting to one orbit $O_{IJ}$ of $g_{IJ}$, reordering the Majoranas gives
\begin{equation}
    \prod_{q\in O_{IJ}} \ii b_{I,g_{IJ}(q)}b_{J,q} =(-1)^{\abs{O_{IJ}}-1} \prod_{q\in O_{IJ}} \ii b_{I,q}b_{J,q},
\end{equation}
while commuting the gauge insertions through the same product costs $(-1)^{\sum_{q\in O_{IJ}}(s_{I,q}+s_{J,q})}$, since each $G$ anticommutes with the dimer bilinear at its site. The term thus equals $(-1)^{\abs{O_{IJ}}-1+\sum_{q\in O_{IJ}}(s_{I,q}+s_{J,q})}$ times itself and vanishes unless the exponent is even. Repeating this argument for every pair $I<J$ (with $g_D=e$) and every dimer flavor~\cite{sheffer2025probing}, a configuration $\{s_{I,r}\}$ contributes only if
\begin{equation}
  \sum_{r\in O}\bigl(s_{I,r}+s_{J,r}\bigr)\equiv \abs{O}-1 \pmod 2
  \label{eq:allowed-sets-rule}
\end{equation}
for every orbit $O$ of $g_{IJ}$.
\end{document}